\documentstyle[12pt,lathuile]{article}
\newcommand{\lsim}{\mathrel{\mathop{\kern 0pt \rlap
  {\raise.2ex\hbox{$<$}}}
  \lower.9ex\hbox{\kern-.190em $\sim$}}}
\newcommand{\gsim}{\mathrel{\mathop{\kern 0pt \rlap
  {\raise.2ex\hbox{$>$}}}
  \lower.9ex\hbox{\kern-.190em $\sim$}}}
\newcommand{\beq}    {\begin{equation}}
\newcommand{\eeq}    {\end{equation}}
\newcommand{\beqarr} {\begin{eqnarray}}
\newcommand{\eeqarr} {\end{eqnarray}}
\newcommand{\barr}   {\begin{array}}
\newcommand{\earr}   {\end{array}}

\begin{document}
\title{ 
PARTICLE CANDIDATES FOR DARK MATTER: A CASE FOR (DOMINANT OR
SUBDOMINANT) RELIC NEUTRALINOS
}
\author{
A. Bottino, N. Fornengo, S. Scopel \\
{\em Dipartimento di Fisica Teorica, Universit\`a di Torino} \\
{\em and INFN, Sez. di Torino, Via Giuria 1, I-10125 Torino, Italy} \\
F. Donato \\
{\em Laboratoire de Physique  Th\'eorique LAPTH, B.P. 110, F--74941}\\
{\em Annecy--le--Vieux Cedex, France} \\
{\em and INFN, Sede di Presidenza, 00186 Roma, Italy} \\
}
\maketitle
\baselineskip=14.5pt
\begin{abstract}
After a short introduction on particle candidates for dark matter
within  possible extensions of the standard
model, we concentrate on Weakly Interacting Massive Particles,
and on one of their most interesting physical realizations: the 
neutralino. 
We analyze how detectability of relic neutralinos by direct and
indirect means is related to their local and cosmological densities;
we use simple general arguments 
to discusss different scenarios where relic neutralinos  make
up the dominant bulk of dark matter or only a  small fraction of it. 
Our general arguments are further corroborated by specific
numerical results. 
We show  to which extent the present experiments of direct searches 
for WIMPs, when interpreted in terms of relic neutralinos,  probe
interesting regions of the supersymmetric parameter space. Our analysis  
is performed in a number of different supersymmetric  schemes.
\end{abstract}
\baselineskip=17pt
\newpage
\section{Introduction}

Evidence for existence  of dark matter and dark energy in the Universe 
is provided by a host of observational data: properties of galactic 
halos and clusters of galaxies, large scale structures, cosmic 
microwave background, high--redshift  supernovae SNe Ia.  
  As far as dark matter is concerned,  a 
favorite range for $\Omega_m h^2$ ($\Omega_m$ being the matter density 
divided by the critical density and $h$ the present--day value of the 
Hubble constant in units of 
${\rm km} \cdot {\rm s}^{-1} \cdot \; {\rm Mpc}^{-1}$) may be set as:   
0.05 $\lsim  \Omega_m h^2  \lsim$ 0.3. 
  Notice that the most recent determinations of cosmological
parameters \cite{bmfr} appear to pin down the matter relic abundance to a
narrower range 0.08 $\lsim \Omega_{m} h^2 \lsim$ 0.21; however, some
caution in taking this range too rigidly is advisable, since some
determinations of cosmological parameters are still subject to fluctuations.
We point out that in the present paper we do not restrict ourselves to any
particular interval of $\Omega_m h^2$; only some features of Figs. \ref{fig:6a}-\ref{fig:6b},
depend on the actual value employed for the minimum amount of matter
necessary to reproduce the halo properties.     

\section{Particle
  candidates for dark matter}

 Various possible extensions of the standard model, envisaged in 
particle physics for reasons quite independent of any cosmological 
motivation, offer,  as an extra bonus, a great variety of particle 
candidates for dark matter. 

A very natural candidate is the 
neutrino: both a light neutrino, with a mass 
$m_{\nu} \lsim 1$ eV, or a heavy one, with a mass 
$m_{\nu} \gsim 100$ GeV, might be  viable and interesting 
candidates.  The light neutrino will be briefly discussed in the 
next section.  

What appears to be the most natural (though certainly not the unique) 
solution of the strong CP--violation, the axion \cite{peccei},  
would also be an appealing candidate for dark matter; 
in fact,  though already constrained by a number of observational 
data, the axion still offers interesting cosmological perspectives. Lack of 
space does not allow here a presentation of the current 
situation; we refer to Ref.\cite{raffelt} for an overview of this subject. 

One of the most favorite candidates for dark matter is the neutralino, 
which in many supersymmetric schemes turns out to be the Lightest 
Supersymmetric Particle (LPS). Supersymmetry   
is known to be motivated 
by a number of strong theoretical arguments, though it is not yet 
supported by experimental evidence, except for a few possible  
hints: unification of coupling constants at the grand unification 
scale \cite{kim}, Higgs LEP events \cite{higgs,el3,ellishiggs}, muon anomalous 
magnetic moment \cite{anomalous,dh,outburst,y} 
(the two latter points will be briefly 
discussed in Sect. 5).  If supersymmetry exists in Nature and the 
R--parity  is conserved, the LSP would be stable. 
Depending on the  susy--breaking mechanism and on the sectors of the 
susy parameter space, the LSP may be one of many options:   sleptons, 
squarks, gluino, gravitino, neutralino, axino (if the axion exists). 
When weakly interacting, the LSP has the prerequisites for being a 
good dark matter candidate.  In the present note we discuss in detail 
the neutralino, whose detection rates may reach the level of current 
experimental sensitivities, both as a dominant or a subdominant 
component of dark matter.  Other possible susy candidates are not
discussed here.

\section{Decoupling of particles from the primordial plasma}

  In the standard scenario of the early Universe, relic particles 
originate from their decoupling from the primordial plasma, while in 
thermic equilibrium; the decoupling of any single species occurred 
when the interaction rate of that species with the thermic plasma 
became smaller than the expansion rate of the Universe (at the 
so--called freeze--out temperature). It is customary to define 
as hot (cold) candidates those particles that are relativistic 
(non relativistic), when they decoupled from the thermal medium. 

A typical case of hot candidates is provided by neutrinos with 
a small mass, $m_{\nu} \lsim$ 1 MeV. Their present--day relic 
abundance turns out to be 
$\Omega_{\nu} h^2 = \sum_{\nu}  m_{\nu} $/(93 eV). Actually, we already 
have some indications of non--vanishing neutrino masses. The 
present experimental data may be formulated in terms of difference in 
squared masses, namely: $\Delta m^2 \sim 10^{-5}- 10^{-4}$ eV$^2$ from
solar neutrinos (if MSW effect is at work) \cite{solar}, 
$\Delta m^2 \sim (2-5) \times 10^{-3}$ eV$^2$ from atmospheric
neutrinos \cite{atmo}, 0.2 eV$^2 \lsim \Delta m^2 \lsim $ 1eV$^2$ 
from the LSND experiment \cite{lsnd}.

Using the data on atmospheric neutrinos, a conservative estimate 
gives:  $\Omega_{\nu} h^2 \gsim  5 \times 10^{-4}$, 
{\it i.e.} a value of relic 
abundance which, although small, is already at the level of the 
average density  of visible matter. Notice that 
 $\Omega_{\nu} h^2$ could be sizeably larger, in case of a degeneracy 
in neutrino masses.  Unfortunately, at present, no experimental device 
is capable of measuring these low--energy neutrinos 
($T_{\nu}$ = 1.95 K). Finally, we recall that light neutrinos cannot 
comprise the whole bulk of dark matter; indeed, should they  
constitute a dominant dark matter component, one would have 
difficulties in generating the present cosmological structures.   

A generic Weakly Interacting Massive Particle (WIMP) would behave 
as a cold dark matter candidate. Its relic abundance may be derived to be 

\begin{equation}
\Omega_{WIMP} h^2 \sim \frac{10^{-37} {\rm cm}^2}{<\sigma_{ann} \; v>}, 
\label{eq:om}
\end{equation}

\noindent
 where 
$\sigma_{ann}$ and $v$ are the WIMP pair--annihilation and the
relative velocity, respectively. $ <\sigma_{ann} \; v>$ denotes the
thermal average of the product $(\sigma_{ann} \cdot v)$ integrated from the freeze--out
temperature to the present--day one. What is remarkable is that for a
number of particle candidates it is conceivable that 
$\Omega_{WIMP} h^2$, as given in Eq. (\ref{eq:om}), may be of order
$0.1 - 1$. Even more noticeable is the fact that for some candidates with 
$\Omega_{WIMP} h^2$ = O(0.1) the detection rates are at the level of
current experimental sensitivities. This is the case of the
neutralino, which will be discussed at some length in the following. 
However, it is obvious that detectability of any given candidate is of
the utmost importance, also in the case of a relic particle which can
only contribute for a fraction of the total dark matter density. We
will examine this point in Sect. 4.3. 

One should be aware that also decoupling mechanisms different from the 
standard one are viable. An interesting example is provided by the
low reheating--temperature scenario, discussed in Ref.\cite{gkr}. 
For instance, in this case the WIMP relic abundance could be proportional
to   $\sigma_{ann}$,   at variance with Eq. (\ref{eq:om}). Thus, the 
relevant WIMP phenomenology could be quite different from the one
depicted in the rest of this note. 

Dark matter particles may also be produced non--thermally (see, for instance,
Ref. \cite{vb} for a review and Ref. \cite{zhang} for a specific
realization leading to relic neutralinos).

\section{Direct and indirect signals of WIMPs}

Different strategies for detecting the presence of 
 WIMPs in our halo have been envisaged. 
The present note is mostly devoted to an analysis of the discovery 
potential of the direct searches for WIMPs, though some considerations 
on the indirect means are also included.

\subsection{WIMP direct detection}

The most natural experimental mean for detecting WIMPs
 is based on the measurement of the recoil that a  nucleus of an 
appropriate detector would suffer, when hit by a WIMP. Various 
 methods may be employed for the detection of the recoil energy 
  \cite{morales}. With the present experimental sensitivities, the 
only signature for disentangling the  signal from the background 
is based on the  annual--modulation effect of the signal, due to the 
composition of the solar-system speed relative to the dark halo with the 
rotational velocity of the Earth around the Sun \cite{freese}. 
As we  show in the following, the discovery potential of 
WIMP direct searches confers to  this experimental mean a prominent  
 role in the investigation of particle dark matter.

       Experiments for WIMP direct detection provide a measurement (or an 
upper bound)  of the differential event rate 

\begin{equation}
\frac {dR}{dE_R}=N_{T}\frac{\rho_{W}}{m_{W}}
                    \int \,d \vec{v}\,f(\vec v)\,v
                    \frac{d\sigma}{dE_{R}}(v,E_{R}),  
\label{eq:diffrate0}
\end{equation}

\noindent
where $N_T$ is the number of the target nuclei per unit of mass, $m_W$
is the WIMP mass, $\rho_W$ is the local WIMP matter density, $\vec v$
and $f(\vec v)$ denote the WIMP velocity and the velocity distribution
function in the Earth frame ($v = |\vec v|$) and $d\sigma/dE_R$ is the
WIMP--nucleus differential cross section.  The nuclear recoil energy, 
$E_R$, 
is given by $E_R={{m_{\rm red}^2}}v^2(1-\cos \theta^*)/{m_N}$, where
$\theta^*$ is the scattering angle in the WIMP--nucleus
center--of--mass frame, $m_N$ is the nuclear mass and $m_{\rm red}$ is
the WIMP--nucleus reduced mass.  Eq.(\ref{eq:diffrate0}) refers to the
case of a monoatomic detector, like the Ge detectors. Its
generalization to more general situations, like for instance the case
of NaI, is straightforward. In what follows, $\rho_W$ will be
factorized in terms of the local value for the total non--baryonic
dark matter density $\rho_l$ and of the fractional amount of density,
$\xi$, contributed by the candidate WIMP, {\it i.e.} $\rho_W = \xi \cdot
\rho_l$. For $\rho_l$ we use the range 0.2 GeV cm$^{-3} \leq \rho_l
\leq$ 0.7 GeV cm$^{-3}$, where the upper side of the range takes into
account the possibility that the matter density distribution is not
spherical, but is described by an oblate spheroidal distribution
\cite{bt,t}.

The WIMP--nucleus differential cross section may conveniently be split
into a coherent part and a spin--dependent one
$\frac {d\sigma}{dE_R} = \left(\frac{d\sigma}{d E_R}\right)_C+
                        \left(\frac{d\sigma}{d E_R}\right)_{SD}$, 
 whose generic features are discussed
in the seminal paper of Ref.\cite{gw}.
 To compare theoretical expectations with experimental data, and
experimental data of different detectors among themselves, it is
useful to convert the WIMP--nucleus cross--section into a
WIMP--nucleon cross section.
Under the hypothesis that the coherent cross--section is dominant and 
 the WIMP couples equally  to protons and neutrons 
(at least approximately), the   WIMP--nucleus cross section
may be expressed in terms of a WIMP--nucleon scalar cross section
 $\sigma^{\rm (nucleon)}_{\rm scalar}$  as  \cite{bdmsbi}
   
\begin{equation}
\frac {d\sigma}{dE_R} \simeq \left ( \frac{d\sigma}{dE_R} \right )_C
              \simeq \frac{F^2(q)}{E^{max}_R}
              \left(\frac{1+m_W/m_p}{1+m_W/m_N}\right )^2 A^2
              \sigma^{\rm (nucleon)}_{\rm scalar}, 
\label{eq:diffrate_approx}
\end{equation}

\noindent where $m_p$ is the proton mass,
$A$ is the nuclear mass number, $E_R^{max}$ is the maximal recoil
energy and $F(q)$ is the nuclear form factor for coherent
interactions, 
  usually parametrized in the Helm form \cite{helm}. In the following 
 we assume that the previous conditions are satisfied, so that, by using
 Eq. (\ref{eq:diffrate_approx}),  a
WIMP--nucleon scalar cross section  $\sigma^{\rm (nucleon)}_{\rm
  scalar}$ 
 may be derived from the WIMP--nucleus
cross section.

From the general formula  in Eq. (\ref{eq:diffrate0}), 
it is clear that, in order to extract  
a WIMP--nucleus cross section from 
the experimental data, one has to use  a specific expression for 
the velocity  distribution function $f(\vec{v})$ (in writing 
  Eq. (\ref{eq:diffrate0}) we have already assumed 
that the WIMP phase--space distribution function is factorizable  as 
$\rho(\vec{r}) \cdot f(\vec{v})$, though this is certainly not the most general case
\cite{bt}).   
  The usual choice for $f(\vec{v})$ is the isotropic  Maxwell--Boltzmann 
distribution in the galactic rest frame, as derived from the 
isothermal-sphere model. The results from  WIMP direct measurements, 
as derived by employing 
the standard isotropic  Maxwell--Boltzmann distribution, 
will be presented in the following section; the
consequences of using other WIMP distributions will be discussed in
Sect. 4.1.2.

\subsubsection{Results from WIMP direct measurements}

\begin{figure}[t]
 \vspace{9.0cm}
\includegraphics{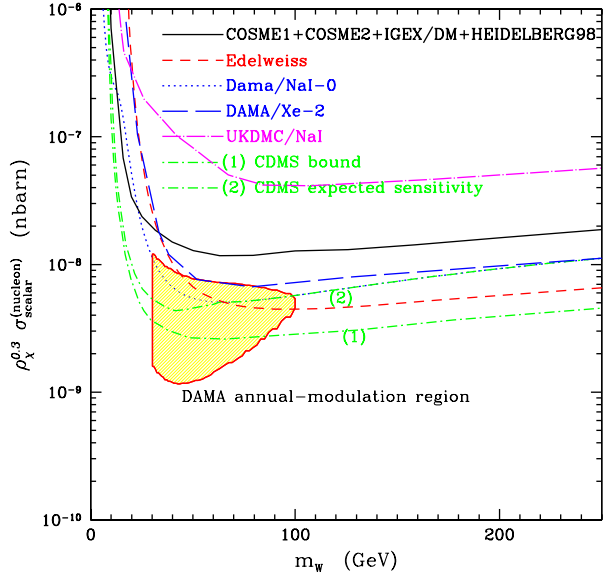}
 \caption{\it
      Experimental results of WIMP direct measurements 
$^{28-35)}$ 
%\cite{ge,edel,dama0,damaxe,boulby,dama,damalast,cdms} 
in terms of  
$\xi \sigma^{\rm (nucleon)}_{\rm scalar}$ as a function of the WIMP mass $m_W$.
The isotropic  Maxwell--Boltzmann distribution has been used for the
  velocity distribution $f(\vec v)$ and the local WIMP density has
  been set at the default value $\rho_l = \;$ 0.3 GeV cm$^{-3}$. 
The open curves denote upper bounds, the closed one denotes the
3--$\sigma$ region, derived from the  annual--modulation 
effect  measured by the DAMA Collaboration \cite{dama,damalast}. 
    \label{fig:1} }
\end{figure}

In Fig. \ref{fig:1} we summarize the most recent experimental results of WIMP direct
measurements
%$^{\rm 23-30)}$ 
\cite{ge,edel,dama0,damaxe,boulby,dama,damalast,cdms} 
in terms of
$\sigma^{\rm (nucleon)}_{\rm scalar}$ as a function of the WIMP mass.  In
deriving the plots of Fig. \ref{fig:1}, the isotropic Maxwell--Boltzmann distribution has
been used for the velocity distribution $f(\vec v)$ and the local WIMP density
has been set at the {\it default} value $\rho_l = \;$ 0.3 GeV cm$^{-3}$.  The
open curves denote upper bounds, the closed one denotes the 3--$\sigma$ region,
derived from the annual--modulation effect measured by the DAMA Collaboration
\cite{dama,damalast}.

 It has   been shown \cite{noi,comp,noi5,noiult,probing,an} that the DAMA 
annual--modulation 
effect  may be  interpreted as due to  relic neutralinos, whose relic 
abundance may also be in the range of  cosmological interest.  
  Comparisons of the
experimental data of Ref.\cite{damalast} with susy calculations have
also been presented in Refs.\cite{kkk,efo,acc,gabr,el2}. 
 In Sect. 5 we will discuss these properties, by proving that  
the current direct experiments for
WIMPs, when interpreted in terms of relic neutralinos, probe 
 regions of the supersymmetric parameter space compatible with
all present bounds from accelerators. 

\subsubsection{Dependence on the WIMP distribution function in the halo}

Let us now discuss some properties related to the use of WIMP distribution
functions different from the standard Maxwell-Boltzmann distribution. 
Recent investigations have shown that deviations from this
standard scheme, either due to a bulk rotation of the dark halo
\cite{kaki,dfs} or to an asymmetry in the WIMP velocity distribution
\cite{vu,ecz,amg}, influence the determination of the WIMP--nucleus
cross sections from the experimental data in a sizeable way.  In
Ref.\cite{ecz} 
also triaxial matter distributions are considered; in the
present paper deviation from sphericity in the WIMP matter
distributions are taken into account only through the physical range
allowed for the value of $\rho_l$ (see our previous comment on
$\rho_W$ after Eq. (\ref{eq:diffrate0})).  In the typical plots of 
$\sigma^{\rm (nucleon)}_{\rm scalar}$ {\it vs} $m_W$, 
 as the ones displayed in Fig. \ref{fig:1},  
the effect introduced by the mentioned deviations from the
Maxwell--Boltzmann is generically to elongate the contours towards
larger values of $m_W$. This is for instance the case for the the
annual--modulation region of the DAMA Collaboration \cite{damalast}.
In Fig. 3 of Ref.\cite{noi5} it is shown that, by implementing the
dark halo with a bulk rotation according to the treatment in
Ref.\cite{dfs}, 
the annual--modulation region moves towards larger values
of the WIMP mass, with an elongation which brings the right--hand
extreme from the value of $\, \sim$ 150 GeV to $\, \sim$ 200 GeV.  A
similar effect is obtained by introducing an asymmetry in the WIMP
velocity distribution $f(\vec{v})$: Fig. 4 of Ref.\cite{amg}
illustrates this point; notice that this asymmetry effect also pushes
somewhat downwards the annual--modulation region.  We emphasize that
all these effects are extremely important, when experimental results
of WIMP direct detection are being compared with theoretical models
for specific candidates.  This point has been overlooked in most
analyses in terms of relic neutralinos.

\subsubsection{WIMP local density}

The values to be assigned to the WIMP local density
$\rho_{W} = \xi \cdot \rho_l$ have to be consistent  with the values of
the WIMP relic abundance, as derived from evaluations for any given 
specific WIMP candidate.  For the case  considered in detail in the
following, the neutralino, we proceed in the following way. The relic
abundance $\Omega_{\chi} h^2$ is evaluated in specific supersymmetric
schemes (see Sect. V). 
When  $\Omega_{\chi} h^2 \ge  (\Omega_m h^2)_{min}$, where 
$(\Omega_m h^2)_{min}$ is the minimum value of $\Omega_m h^2$ 
compatible with halo properties, we simply set $\rho_{\chi} = \rho_l$
(i.e., $\xi = 1$). 
 When  $\Omega_{\chi} h^2 <  (\Omega_m h^2)_{min}$,  the neutralino
 cannot be the  unique cold dark matter particle, thus 
we assign to the neutralino a {\it rescaled} local density 
$\rho_{\chi} = \rho_l \times \Omega_{\chi} h^2/(\Omega_m h^2)_{min}$ 
(i.e., $\xi = \Omega_{\chi} h^2/(\Omega_m h^2)_{min}$) \cite{gst}. 
Thus, summarizing,

\begin{eqnarray} 
&\rho_{\chi}& = \rho_l, \; \;  \; \; \; \; \; 
{\rm when} \; \; \Omega_{\chi} h^2 \ge (\Omega_m h^2)_{min} 
\label{eq:rescaling1} \\ 
&\rho_{\chi}& = \frac{\Omega_{\chi} h^2}{(\Omega_m h^2)_{min}} \; \rho_l, 
\;  \; \; \;  
{\rm when} \; \; \Omega_{\chi} h^2 < (\Omega_m h^2)_{min} 
\label{eq:rescaling2}.
\end{eqnarray}

  We stress that in our analyses we consider various scenarios, where
  neutralinos may constitute the dominant cold dark matter component or only a
   small fraction of it. 
As discussed below (see Sect. 4.3), in view  of the properties
related to detectability of relic neutralinos, it is very important 
not to disregard neutralino configurations with small relic
abundance, where rescaling applies.

\subsubsection{Sensitivity range for current WIMP direct searches}

In the present paper we focus  our analysis to the WIMP mass
range which, in the light of the experimental data summarized
above and of the previous considerations on the
astrophysical uncertainties, appears
particularly appealing:  

\begin{equation}
40 \; {\rm GeV} \leq  m_W \leq 200 \;  {\rm GeV}. 
\label{eq:mass}
\end{equation}

Notice that the mass range of Eq. (\ref{eq:mass}) is 
quite appropriate for neutralinos. Actually, the lower extreme 
is indicative of the LEP lower bound on the neutralino mass 
$m_{\chi}$ (in the calculations performed in the present work the
actual lower bound for $m_{\chi}$, dependent on the other susy 
parameters, is employed, according to the
constraints given in \cite{LEPb}). 
As for the upper 
extreme, we notice that, though a generic  range for $m_{\chi}$
 might extend up to about 1 TeV, requirements of no excessive 
fine--tuning  \cite{bere1}
would actually  favour an upper bound of order 200 GeV, 
in accordance with Eq. (\ref{eq:mass}).  

In what follows we will discuss the discovery potential of WIMP direct
searches for WIMPs in the mass range of Eq.(\ref{eq:mass}). Particular
attention will be paid to capabilities of the present experiments; 
according to the discussion previously made, 
their sensitivity range, in case of WIMPs whose coherent interactions
with ordinary matter are dominant over the the spin--dependent ones,
may be stated, in terms of the quantity $\xi \sigma^{\rm
  (nucleon)}_{\rm scalar}$, as

\begin{equation}
4 \cdot 10^{-10} \; {\rm nbarn} \leq \
\xi \sigma^{\rm (nucleon)}_{\rm scalar} \leq 
 2 \cdot 10^{-8} \; {\rm nbarn}.
\label{eq:section}
\end{equation}

We will hereafter refer to region $R$ as the one in the space 
$m_W - \xi \sigma^{\rm (nucleon)}_{\rm scalar}$ which is  defined by 
Eqs. (\ref{eq:mass}-\ref{eq:section}). The region $R$ represents the
sensitivity region already under exploration with present
detectors.

Our analysis, based on an interpretation of experimental data in
terms of relic neutralinos,  will show by how much the
WIMP direct searches probe the supersymmetric parameter space.

\subsection{WIMP indirect searches}

Indirect searches for WIMPs aim at the measurements of signals which 
fall essentially into two categories (for
a review and relevant references see, for instance, Ref.\cite{bf}): 

i) Signals due to WIMP pair--annihilations taking place in our
galactic halo. The most interesting outputs of these annihilation
processes are: a) fluxes of neutrinos and gammas (which might be
disentangled from the background in case of halo clumpiness), b) a
narrow gamma-gamma line, c) exotic components in cosmic rays:
antiprotons, positrons, antideuterons. 

ii) Signals due to WIMP pair--annihilations taking place in the
interior of celestial bodies.   
These signals would consist of fluxes of up--going muons in a 
neutrino telescope, generated by neutrinos which are produced by the 
pair annihilations of WIMPs captured and accumulated inside the 
Earth and the Sun. 

For the discussion to follow, it is important to notice that in case
(i) the detection rates are proportional to 
$\rho_{W}^2 \cdot \sigma_{ann}$, whereas in case (ii) the detection
rates are proportional to  
$\rho_{W} \cdot \sigma^{\rm (nucleon)}_{\rm scalar}$ (in fact, in
this latter case, the annihilation rate inside the celestial body is
proportional to the capture rate of the WIMPs by the body, which is in
turn 
proportional to $\rho_{W} \cdot \sigma^{\rm (nucleon)}_{\rm scalar}$).

\subsection{General arguments on detectability of WIMPS versus their  
local density and average relic abundance}

One particular type of relic particle (say, a WIMP) may constitute 
a dominant or a subdominant dark matter candidate. Also the latter
case would be of great physical interest, provided this relic particle
has some chance to be detected. Thus, a crucial question is: 
How do experimental detectablities of WIMPs depend on the WIMP 
(local and cosmological) densities ?

This question may be answered by simple general arguments which 
we already presented in a number of papers 
\cite{noi94,kk,bere1,anti,comb,noi6,probing}
 and which we briefly review here. 
Our arguments are exemplified by properties of relic neutralinos, 
in case these particles decouple from the primordial plasma when
 in thermodynamic 
equilibrium. It is obvious that these same arguments apply equally well 
to other realizations of WIMPs, under the same decoupling conditions. 

We consider the rates of direct signals and  of signals  
due to pair annihilation in celestial bodies, separately from the rates  
of signals due to  neutralino--neutralino annihilations
taking place in the galactic halo, because of their different
dependence  on local density and cross sections. 

\subsubsection{Rates of direct detection and of signals  
due to pair annihilation in celestial bodies}

 As was seen  above, 
in both of these experimental measurements  
 the detection rate $R$ is proportional to the product 
$\rho_{\chi} \cdot \sigma_{\rm scalar}^{(\rm nucleon)}$.  Thus, taking into 
account the rescaling properties of $\rho_{\chi}$
(Eqs. (\ref{eq:rescaling1})-(\ref{eq:rescaling2})), we have that 
$R$ behaves as follows

\begin{eqnarray} 
&R& \; \; \propto \; \sigma_{\rm scalar}^{(\rm nucleon)}, \; \;  \; \; \;
{\rm when} \; \; \Omega_{\chi} h^2 \ge (\Omega_m h^2)_{min} \label{eq:1} \\ 
&R& \; \; \propto \; \frac{\Omega_{\chi} h^2}{(\Omega_m h^2)_{min}} \; 
\sigma_{\rm scalar}^{(\rm nucleon)} \propto 
\frac{\sigma_{\rm scalar}^{(\rm nucleon)}}{<
  \sigma_{ann}v >}, 
\;  \; \; \;  
{\rm when} \; \; \Omega_{\chi} h^2 < (\Omega_m h^2)_{min} \label{eq:2}.
\end{eqnarray}

\begin{figure}[t]
 \vspace{9.0cm}
\includegraphics{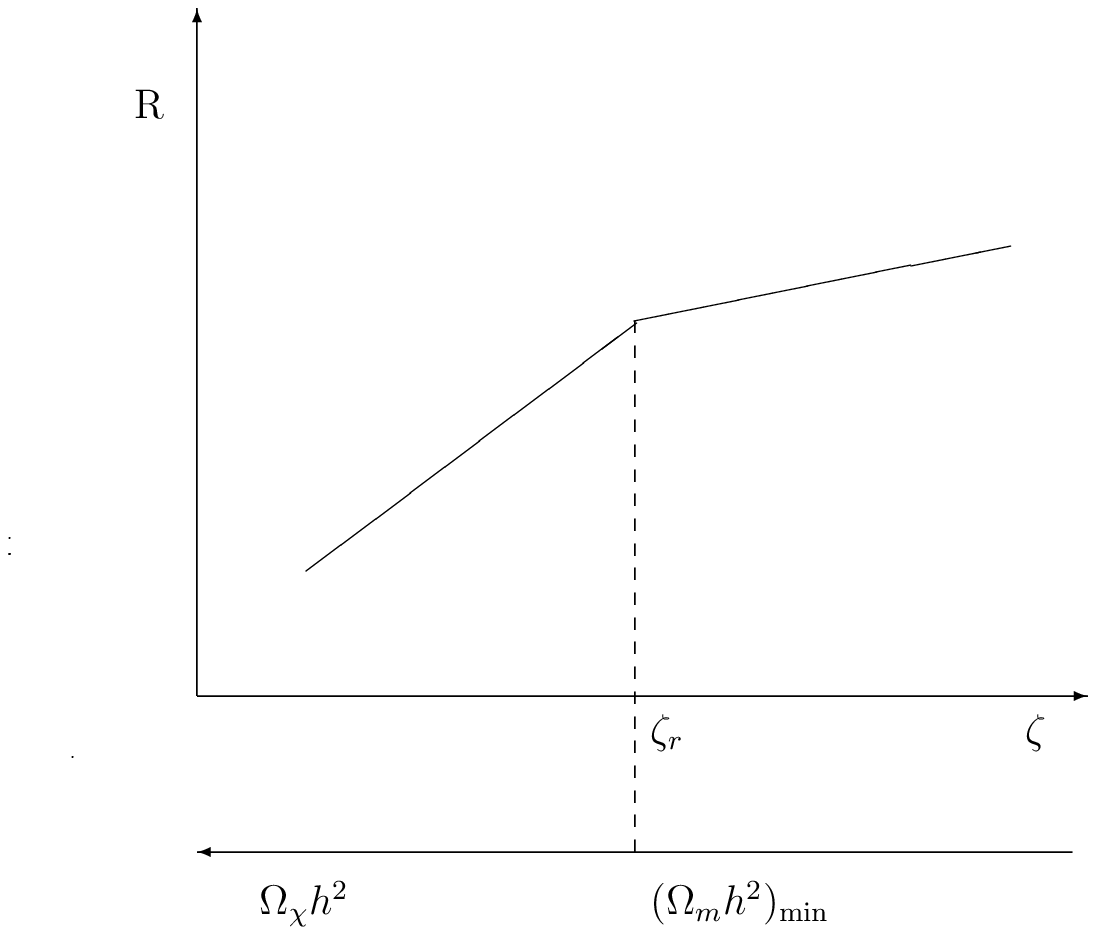}
 \caption{\it
Qualitative behaviour of the rates of direct detection and of signals  
due to pair annihilation in celestial bodies as functions of a generic 
coupling parameter $\zeta$ ($\zeta_r$  denotes the value of $\zeta$ at which 
$\Omega_{\chi} h^2 = (\Omega_m h^2)_{min}$). Notice the correlation
between the rates $R$ and the neutralino relic abundance 
$\Omega_{\chi} h^2$.
\label{fig:2}
}
\end{figure}

Now, the cross sections $\sigma_{\rm scalar}^{(\rm nucleon)}$ and
$\sigma_{ann}$, as functions of any generic coupling parameter $\zeta$, behave
similarly ({\it i.e.} they usually both decrease or increase in terms of
variations of that parameter), because of crossing symmetry.  Thus, for
instance, in supersymmetric schemes where the neutralino is the LPS,
$\sigma_{\rm scalar}^{(\rm nucleon)}$ and $\sigma_{ann}$ are both increasing
functions of $\tan \beta$, when the relevant processes are mediated by Higgs
bosons.  Usually, $\sigma_{\rm scalar}^{(\rm nucleon)}$ increases somewhat
faster than $\sigma_{ann}$, or approximately at the same rate.  Since, at the
same time, $\Omega_{\chi} h^2 \propto (<\sigma_{ann} v>)^{-1}$, the typical
behaviour of the rate $R$ (for the processes under study) is the one displayed
in Fig. \ref{fig:2}. For small values of $\zeta$, both $\sigma_{\rm scalar}^{(\rm
  nucleon)}$ and $\sigma_{ann}$ are small (and then $\Omega_{\chi} h^2$ is
large) and $R$ grows proportionally to $\sigma_{\rm scalar}^{(\rm nucleon)}$.
As the strength of the coupling increases, in the region beyond the value
$\zeta_r$ (at which $\Omega_{\chi} h^2 = (\Omega_m h^2)_{min}$), Eq.
(\ref{eq:2}) applies: the rate $R$ still increases (though less rapidly than
before rescaling), or remains approximately flat. Therefore, as far as direct
detection and indirect detection through pair annihilation in celestial bodies
are concerned, we obtain that {\it the detectability of relic neutralinos is
  usually favoured for neutralinos of small $\Omega_{\chi} h^2$, that is for
  neutralinos which comprise only a subdominant dark matter component}.

\begin{figure}[t]
 \vspace{9.0cm}
\includegraphics{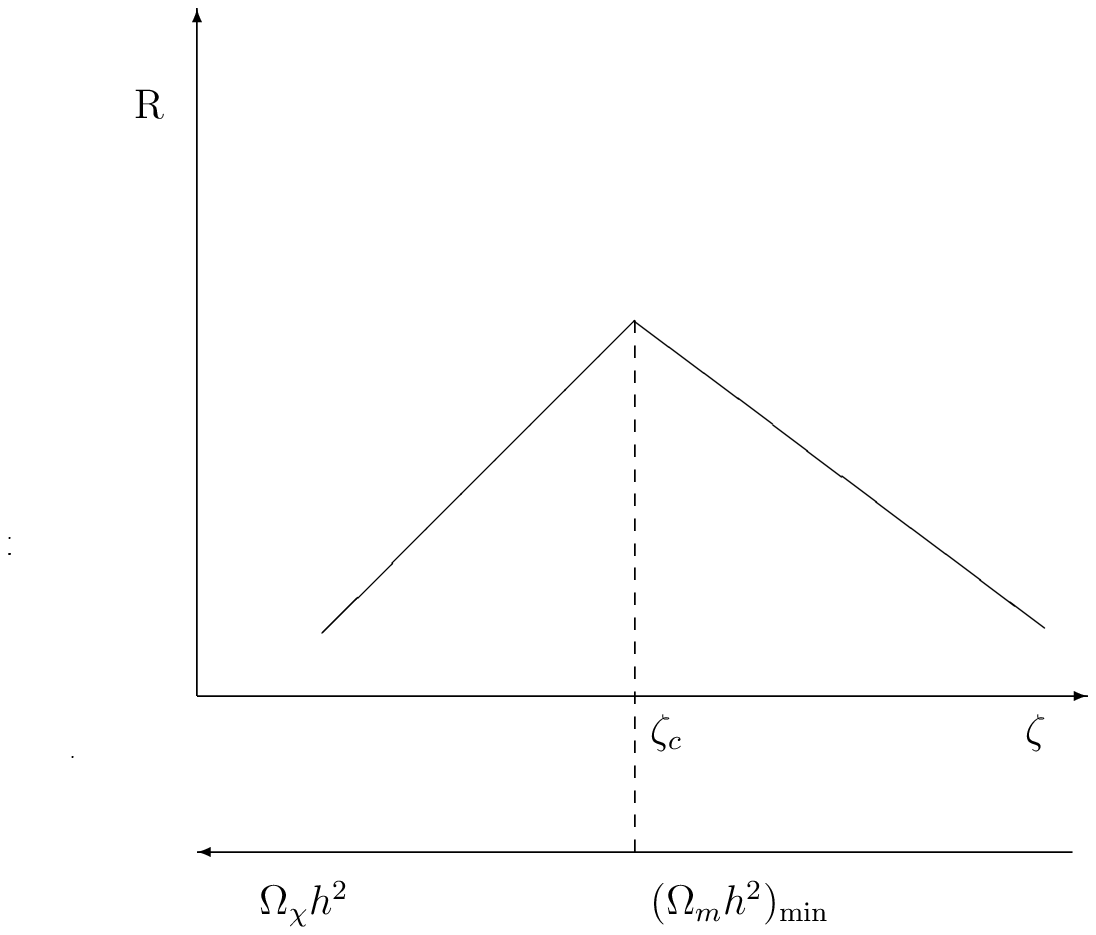}
 \caption{\it
Qualitative behaviour of the rates of signals due to 
neutralino--neutralino annihilations taking place 
in the galactic halo as functions of a generic 
coupling parameter $\zeta$ ($\zeta_r$  denotes the value of $\zeta$ at which 
$\Omega_{\chi} h^2 = (\Omega_m h^2)_{min}$). Notice the correlation
between the rates $R$ and the neutralino relic abundance 
$\Omega_{\chi} h^2$.
\label{fig:3}
}
\end{figure}

We stress that, due to this last property, in the numerical analyses 
of the properties of relic neutralinos, it is important   
to keep into consideration also susy configurations where rescaling 
applies. In fact, by neglecting these configurations, one would 
disregard relic 
neutralinos which have the largest chances of being detected. 
 Because of this fact, at variance with most 
analyses by other authors, our scanning of the susy 
parameter space always included  configurations entailing 
small neutralino relic abundances (see, for instance, 
\cite{probing}).

The numerical evaluations, which we present in the next section, show 
that the actual situation is even better; in fact, it turns out that 
{\it  a   number of  neutralino configurations of  cosmological interest,  
though disfavoured by the previous arguments,
may reach the level of detectability by current experiments}. Actually,
we will prove that the annual--modulation effect, measured by the DAMA
Collaboration, is compatible with an interpretation in terms of
neutralinos, part of which with relic abundance 
of cosmological interest. Due to the 
connection between $\sigma_{\rm scalar}^{(\rm nucleon)}$ and 
$\Omega_{\chi} h^2$, this property is far from being trivial.

\subsubsection{Rates of signals due to 
neutralino--neutralino annihilations taking place 
in the galactic halo.}
 
At variance with the previous case, the 
detection rates have now a {\it quadratic} dependence on the neutralino 
density in the halo, $R \propto \rho_{\chi}^2 \cdot \sigma_{ann}$; thus, by 
using 
Eqs. (\ref{eq:rescaling1})-(\ref{eq:rescaling2}), 
 we have 
 the following behaviour     for $R$

\begin{eqnarray} 
&R& \; \; \propto \; \sigma_{ann}, \; \;  \; \; \; \; \; 
{\rm when} \; \; \Omega_{\chi} h^2 \ge (\Omega_m h^2)_{min} \label{eq:3} \\ 
&R& \; \;  \propto 
(\frac{\Omega_{\chi} h^2}{(\Omega_m h^2)_{min}})^2 \; \sigma_{ann} \;
\propto \; \frac{\sigma_{ann}}{{< \sigma_{ann}v >}^2}, 
\;  \; \; \;  
{\rm when} \; \; \Omega_{\chi} h^2 < (\Omega_m h^2)_{min} \label{eq:4}.
\end{eqnarray}

Therefore, in  this case, as the generic coupling parameter $\zeta$ 
increases, the detection rate first increases, as long as the couplings are
small; afterwards, for values of $\zeta$
larger than $\zeta_r$, when  rescaling applies, 
the detection rate decreases (see Fig. \ref{fig:3}).

It follows that for processes depending on pair--annihilation in the halo 
the maximal rates occur for values of the relic abundance around 
the value $(\Omega_m h^2)_{min}$. Subdominant neutralinos are 
disfavoured for detectability by this type of signals as compared to 
neutralinos with a relic abundance  around the value  
$\Omega_{\chi} h^2 \simeq (\Omega_m h^2)_{min}$.

  Finally,  we notice that all the arguments  reviewed in this section
 were already presented in our previous papers (see, for instance, 
\cite{noi94,kk,bere1}),  further corroborated by extensive 
numerical evaluations \cite{anti,comb,noi6,probing}. 
Amazingly, in a recent preprint \cite{dgg}, where  
 considerations similar to ours are repeated, all previous references  
have been overlooked.

\section{Supersymmetric dark matter}

We turn now to calculations performed in specific supersymmetric
schemes, assuming that R-parity is conserved, and thus that the 
LSP is stable. The nature of the LSP depends on the susy--breaking
mechanism and on the specific regions of the susy parameter space. We consider
here gravity--mediated schemes, and domains of the parameter space, 
where the LSP is the neutralino. Extensive calculations on relic
neutralino phenomenology in gravity--mediated models 
have been performed (see, for instance, 
Refs.$^{{\rm 31-41,49,51-55,57,59-65)}}$).

%\cite{noi,comp,noi5,noiult,probing,an,kkk,efo,acc,gabr,
%el2,noi94,kk,bere1,anti,comb,noi6,cn,fmw,nnn,noiom,bg,man,lns,altri}). 

Our analysis will follow 
the one reported in Ref.\cite{probing} and will show 
by how much the
WIMP direct searches probe the supersymmetric parameter space. 
We remark  that, in the case of neutralinos, the assumption about the 
dominance  of the coherent cross section over the spin--dependent one 
is, in general, largely satisfied, except for values of 
 $\sigma^{\rm (nucleon)}_{\rm scalar}$ which are 
far below  the present   experimental reach \cite{bdmsbi}.
 
\subsection{Susy models}

The analysis of Ref.\cite{probing} was performed in  the Minimal
Supersymmetric extension of the Standard Model (MSSM) in a variety of
different schemes,  from those based on universal or non-universal 
supergravity, with susy parameters defined at the grand unification
scale, to an
effective supersymmetric model defined at the Electro--Weak (EW)
scale. Here we only report about results in universal supergravity and
in  the effective scheme at EW scale; for the other cases and further
details we refer to Ref.\cite{probing}.  

  The essential elements of the MSSM are described
by a Yang--Mills Lagrangian, the superpotential, which contains all
the Yukawa interactions between the standard and supersymmetric
fields, and by the soft--breaking Lagrangian, which models the
breaking of supersymmetry.  The
Yukawa interactions are described by the parameters $h$, which
are related to the masses of the standard fermions by the usual
expressions, {\em e.g.}, $m_t = h^t v_2, m_b = h^b v_1$, where $v_i$
are the $vev$'s of the two Higgs fields, $H_1$ and $H_2$.
 Implementation of  this model within a supergravity scheme 
 leads naturally to a set of unification assumptions at a Grand
 Unification (GUT) scale, $M_{GUT}$:

     i) Unification  of the gaugino masses:
        $M_i(M_{GUT}) \equiv m_{1/2}$,

     ii) Universality of the scalar masses with a common mass denoted by
     $m_0$: $m_i(M_{GUT})$ \hfill \break
    \indent \phantom{ii)\ }  $ \equiv m_0$,

    iii) Universality of the trilinear scalar couplings:
         $A^{l}(M_{GUT}) = A^{d}(M_{GUT}) = A^{u}(M_{GUT})$ \hfill \break
    \indent \phantom{iii)\ }  $\equiv A_0 m_0$. 
    
    This scheme will be denoted here as universal SUGRA (or simply SUGRA). The
    relevant parameters of the model at the electro--weak (EW) scale are
    obtained from their corresponding values at the $M_{GUT}$ scale by running
    these down according to the renormalization group equations (RGE). By
    requiring that the electroweak symmetry breaking is induced radiatively by
    the soft supersymmetry breaking, one finally reduces the model parameters
    to five: $m_{1/2}, m_0, A_0, \tan \beta (\equiv v_2/v_1)$ and sign $\mu$.
    Here the parameters are varied in the following ranges: $50\;\mbox{GeV}
    \leq m_{1/2} \leq 1\;\mbox{TeV},\; m_0 \leq 1\;\mbox{TeV},\; -3 \leq A_0
    \leq +3,\; 1 \leq \tan \beta \leq 50$.  Notice that a common upper extreme
    for the mass parameters has been used, and generically set at the value of
    1 TeV, as a typical scale beyond which the main attractive features of
    supersymmetry fade away. However, fine-tuning arguments actually set
    different bounds for $m_0$ and $m_{1/2}$ (in universal SUGRA and in
    non--universal SUGRA) \cite{bere1}: $m_{1/2} \lsim$ hundreds of GeV, whereas $m_0 \lsim
    2-3$ TeV. The $m_0 \sim 2-3$ TeV window, which is not specifically included
    in the results reported here, will be analysed in a forthcoming paper
    \cite{ourfuture}. In Ref.\cite{fmw} properties of relic neutralinos in this
    large $m_0$ regime have been analyzed. We remark that the phenomenology of
    relic neutralinos is also very sensitive to other parameters, such as the
    top quark mass $m_t$ and the strong coupling $\alpha_s$. For these
    parameters, we use here their 95\% CL ranges: $m_t = (175 \pm 10)$ GeV and
    $\alpha_s(M_Z) = 0.118 \pm 0.004$.

Models with unification conditions at the GUT scale
represent an  appealing scenario; however,
some of the assumptions listed above, particularly ii) and iii), are not
very solid, since, as was  already emphasized some time ago \cite{com},
universality might occur at a scale higher than $M_{GUT}\sim 10^{16}$
GeV, {\em e.g.}, at the Planck scale. More recently, the possibility that
 the initial scale for the RGE running, $M_I$, might be smaller than 
 $M_{GUT}\sim 10^{16}$ has been raised \cite{gabr,abel}, on the basis of
 a number of string models (see for instance the references quoted in 
\cite{gabr}).  In
Ref.\cite{gabr} it is stressed that $M_I$ might be anywhere between the EW
scale and the Planck scale, with significant consequences  for the size of
the neutralino--nucleon cross section.  

An empirical way of taking into account the uncertainty in $M_I$ 
 consists in allowing deviations in the
unification conditions at $M_{GUT}$.     For instance, deviations from 
universality in the scalar  masses at  $M_{GUT}$, which split 
$M_{H_1}$ from $M_{H_2}$ may be parametrized as

\begin{equation}
M_{H_i}^2 (M_{GUT}) = m_0^2(1 + \delta_i)
\label{eq:dev}
\end{equation}

This is the case of non--universal SUGRA (nuSUGRA) that we considered 
 in Refs. \cite{comp,bere1,probing}. 
Further extensions of deviations from universality in SUGRA models
 which  include squark and/or gaugino masses are discussed, for instance,
 in \cite{acc,cn}. In this note we do not report results in nuSUGRA
 schemes; we refer to Ref.\cite{probing} for discussions on this
 case.

   The large uncertainties involved in the choice of the scale $M_I$ 
make the SUGRA schemes somewhat problematic: 
the originally appealing feature of a universal SUGRA with few parameters 
fails, because of the need to take into consideration the variability of $M_I$ 
or, alternatively, to add new parameters which quantify the various 
deviation effects from universality at the GUT scale. It
appears  more convenient  to work  with a phenomenological 
susy model whose  parameters are defined directly at the electroweak 
scale. We denote here this effective scheme of MSSM by effMSSM. This   
 provides, at the EW scale, a model, defined in terms of a minimum  number of 
parameters: only those necessary to shape the essentials of the theoretical 
structure of an MSSM, and of its particle content. 
Once all experimental and theoretical constraints are implemented in
this effMSSM model, one may investigate its compatibility with SUGRA
schemes at the desired $M_I$.

In the effMSSM scheme we consider here, we impose  a set of
assumptions at the electroweak scale: 
a) all trilinear parameters are set to zero except those of the third family, 
which are unified to a common value $A$;
b) all squark  soft--mass parameters are taken  
degenerate: $m_{\tilde q_i} \equiv m_{\tilde q}$; 
c) all slepton  soft--mass parameters are taken  
degenerate: $m_{\tilde l_i} \equiv m_{\tilde l}$; 
d) the $U(1)$ and $SU(2)$ gaugino masses, $M_1$ and $M_2$, are 
assumed to be linked by the usual relation 
$M_1= (5/3) \tan^2 \theta_W M_2$ (this is the only GUT--induced
relation we are using, since gaugino mass unification appears to be
better motivated than scalar masses universality). 
As a consequence, the supersymmetric 
parameter space consists of seven independent parameters. 
We choose them to be: 
$M_2, \mu, \tan\beta, m_A, m_{\tilde q}, m_{\tilde l}, A$ and vary these 
parameters in
the following ranges: $50\;\mbox{GeV} \leq M_2 \leq  1\;\mbox{TeV},\;
50\;\mbox{GeV} \leq |\mu| \leq  1\;\mbox{\rm TeV},\;
80\;\mbox{GeV} \leq m_A \leq  1\;\mbox{TeV},\; 
100\;\mbox{GeV} \leq  m_{\tilde q}, m_{\tilde l} \leq  1\;\mbox{TeV},\;
-3 \leq A \leq +3,\; 1 \leq \tan \beta \leq 50$ ($m_A$ is the mass of
the CP-odd neutral Higgs boson).

The effMSSM scheme proves  very manageable for the susy phenomenology at the 
EW scale; as such, it has been frequently used in the literature in 
connection with relic neutralinos 
(often with the further assumption of slepton/squark mass
degeneracy: 
$m_{\tilde{q}} = m_{\tilde{l}}$) \cite{noi,noiult,kkk,nnn,bg,man}. 
Notice that we are not assuming here slepton/squark mass degeneracy.

We recall that even much larger extensions of the
supersymmetric models could be envisaged: for
instance,   non--unification of the gaugino masses \cite{cn,griest},
and schemes with CP--violating phases \cite{cp}. 
Here we limit our considerations to the two schemes previously defined: 
universal SUGRA and effMSSM; for nuSUGRA we refer to
Ref.\cite{probing}. 

 The neutralino is defined 
as the lowest--mass linear superposition of photino ($\tilde \gamma$),
zino ($\tilde Z$) and the two higgsino states
($\tilde H_1^{\circ}$, $\tilde H_2^{\circ}$):
$\chi \equiv a_1 \tilde \gamma + a_2 \tilde Z + a_3 \tilde H_1^{\circ}  
+ a_4 \tilde H_2^{\circ}$. 
Hereafter, the nature of the neutralino is classified in terms of a
parameter $P$, defined as $P \equiv a_1^2 + a_2^2$.  
The neutralino is called a gaugino when $P > 0.9$, a higgsino when 
$P < 0.1$, mixed otherwise. 

For more details concerning theoretical aspects involved in our calculations 
 we refer to Refs.\cite{comp,noiult,probing}. 
 Accelerators data on supersymmetric
and Higgs boson searches (CERN $e^+ e^-$ collider LEP2 and Collider
Detector CDF at Fermilab) provide now rather stringent bounds on
supersymmetric parameters. 
CDF bounds are taken from \cite{cdf}. The new LEP2 bounds are taken
from \cite{LEPb,donan}. 

As compared to the calculations presented in Ref.\cite{probing}, we have now
implemented effects due to those radiative corrections to the couplings of the
neutral Higgs bosons to the quarks which may be sizeable at large $\tan \beta$
\cite{qcd_corr}. These radiative corrections affect the calculation of the
neutralino--nucleus cross section and of the neutralino cosmological relic
abundance.  Notice that the correction to the relation between the $b$ quark
mass and its Yukawa coupling, implied by these radiative corrections, enters
also in the calculations of the $b \rightarrow s + \gamma$ decay
\cite{bsg_carena}. For the SUGRA model discussed above, it affects also the
boundary conditions at the GUT scale for the $b$ Yukawa coupling \cite{sarid}.
This in turn affects the radiative symmetry breaking mechanism and the
low--energy Higgs and sfermion spectra \cite{bere1}. All these effects are
included in our calculations.

Finally, we notice that a new experimental constraint on supersymmetric 
 parameters may also  be   derived from the  recent  accurate experimental
determination of the muon anomalous magnetic moment  
\cite{anomalous}; this measurement provides the value 
$a_{\mu}({\rm exp}) = 11 659 202(14)(6) \times 10^{-10} (1.3~{\rm ppm})$, 
where $a_{\mu} = (g - 2)/2$. This data, if compared with the 
theoretical evaluations in Ref.\cite{dh}, would show a deviation 
of 2.6 $\sigma$ from the standard--model prediction. 
  This has determined an outburst of theoretical papers 
\cite{outburst}, where 
this possible deviation is attributed to supersymmetry, and the 
relevant implications derived.  However,
other more recent standard--model evaluations of 
$a_{\mu}$ \cite{n,j} are in better agreement with the experimental data 
of Ref.\cite{anomalous} (see also Refs. \cite{mlr,y} for detailed
 discussions
of the standard--model calculations of $a_{\mu}$). Thus,
for the time being, it appears safer to use the data of Ref. \cite{anomalous} as a constraint
on susy, rather than a sign of it. Employing the theoretical
results of Refs. \cite{dh,n,j}, the
contribution of supersymmetry to the anomalous moment is constrained
by  $-140 \leq a_{\mu}^{susy} \cdot 10^{11} \leq 890$ .  This constraint
has been implemented in our present scanning of the supersymmetric parameter
space. 

\subsection{Numerical results}

\begin{figure}[t]
 \vspace{9.0cm}
\includegraphics{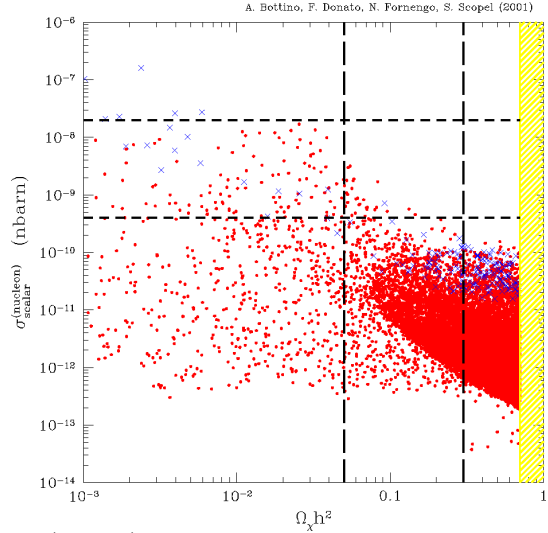}
 \caption{\it
Scatter plot of $\sigma_{\rm scalar}^{(\rm nucleon)}$ versus 
$\Omega_{\chi} h^2$ for  universal SUGRA. Set 1 for the 
quantities $m_{q}<\bar{q}q>$'s is employed \cite{noi6}.
Only configurations with positive $\mu$ are shown and 
$m_{\chi}$ is taken in the range of Eq. (\ref{eq:mass}).
 The two horizontal lines bracket the sensitivity region defined 
 by Eq. (\ref{eq:section}), for $\xi = 1$. 
 The two vertical lines denote the range 
$0.05 \leq \Omega_{m} h^2 \leq 0.3$.
The region above $\Omega_{\chi} h^2 = 0.7$ is excluded by current limits on
the age of the Universe. 
Dots (crosses)
denote gaugino (mixed) configurations. 
\label{fig:4a}}
\end{figure}

\begin{figure}[t]
 \vspace{9.0cm}
\includegraphics{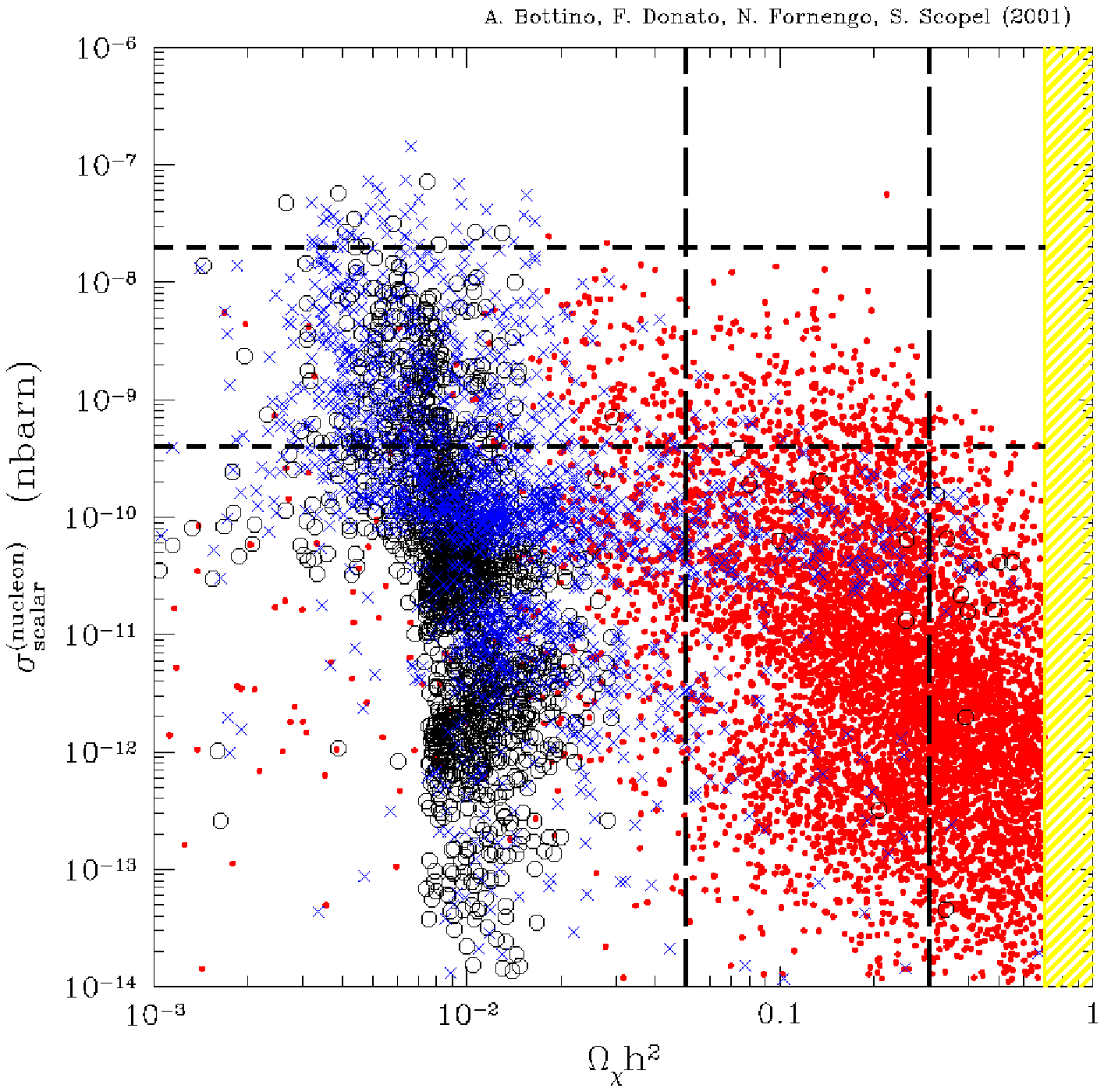}
 \caption{\it
Scatter plot of $\sigma_{\rm scalar}^{(\rm nucleon)}$ versus 
$\Omega_{\chi} h^2$ for  effMSSM. Notations as in Fig. \ref{fig:4a}. 
Dots denote gauginos, circles denote higgsinos and crosses denote mixed 
configurations.  Both signs
of $\mu$ are shown. 
\label{fig:4b}}
\end{figure}

We turn now to the presentation of our results. In Figs. \ref{fig:4a}-\ref{fig:4b} we
give the scatter plots for $\sigma_{\rm scalar}^{(\rm nucleon)}$ versus 
$\Omega_{\chi} h^2$ for two different schemes: universal SUGRA,  and effMSSM. 
For the SUGRA scheme we only display the results corresponding to positive 
values of $\mu$, since, for  negative values,
the constraint on $b \rightarrow s + \gamma$ implies a large suppression of 
$\sigma_{\rm scalar}^{(\rm nucleon)}$. 
  The calculations of
$\sigma^{\rm (nucleon)}_{\rm scalar}$ have been performed with the formulae
reported in Refs.\cite{noi6}; set 1 for the quantities
$m_{q}<\bar{q}q>$'s has been used (see Ref.\cite{noi6} for definitions); the
evaluation of $\Omega_{\chi} h^2$ follows the procedure given in \cite{noiom}.
   The two horizontal lines bracket
the sensitivity region defined by Eq. (\ref{eq:section}), when $\xi$
is set equal to one. The two
vertical lines denote the favorite range for $\Omega_{m}h^2$, 
$0.05 \leq \Omega_{m} h^2 \leq 0.3$.

Figs. \ref{fig:4a}-\ref{fig:4b} provide a first relevant result of our analysis: the present
experimental sensitivity in WIMP direct searches allows the exploration of
supersymmetric configurations compatible with current accelerator bounds. A
number of configurations stay inside the region of cosmological interest, also
in the constrained SUGRA scheme. The region of experimental sensitivity and
cosmological interest is covered with an increasingly larger variety of
supersymmetric configurations as one moves from SUGRA to effMSSM; this latter
fact is expected from the intrinsic features of the various schemes.

\begin{figure}[t]
 \vspace{9.0cm}
\includegraphics{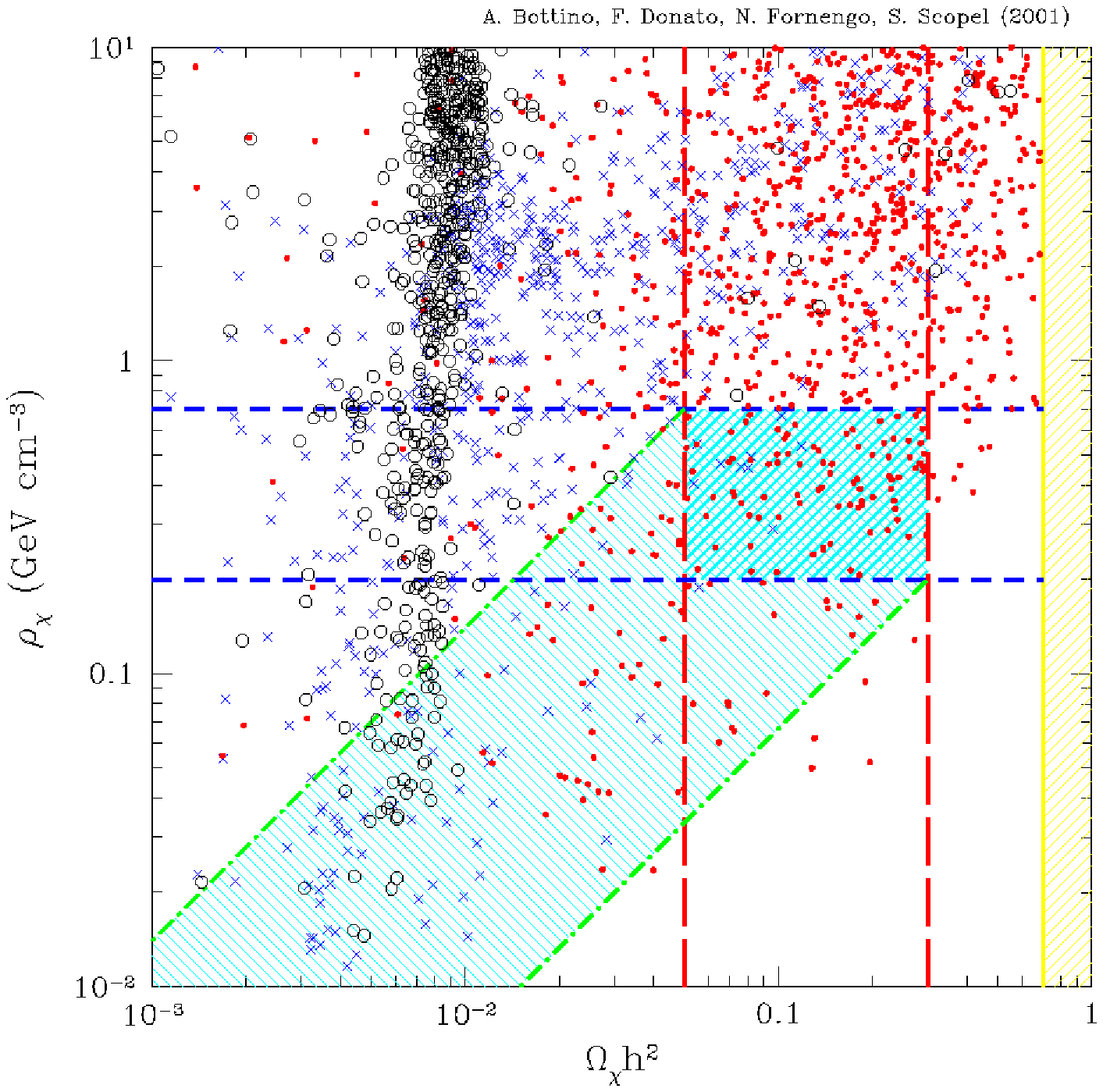}
 \caption{\it
  Scatter plot of $\rho_{\chi}$ versus $\Omega_{\chi}h^2$ for the
  effMSSM.  This plot is derived from the experimental value
  $[\rho_{\chi}$/(0.3 GeV cm$^{-3}$) $\cdot \sigma^{\rm (nucleon)}_{\rm
    scalar}]_{expt} = 1 \cdot 10^{-9}$ nbarn and by taking $m_{\chi}$ in the
  range of Eq. (\ref{eq:mass}), according to the procedure outlined in the
  text.  Set 1 for the quantities $m_{q}<\bar{q}q>$'s is employed \cite{noi6}.  The two
  horizontal lines delimit the range 0.2 GeV cm$^{-3} \leq \rho_{\chi} \leq $
  0.7 GeV cm$^{-3}$; the two vertical ones delimit the range $0.05 \leq
  \Omega_{m} h^2 \leq 0.3$.  The region above $\Omega_{\chi} h^2 = 0.7$ is
  excluded by current limits on the age of the Universe.  The band delimited by
  the two slanted dot--dashed lines and simply hatched is the region where
  rescaling of $\rho_l$ applies.  Dots denote gauginos, circles denote
  higgsinos and crosses denote mixed configurations.
  \label{fig:5}}
\end{figure}

  Once a measurement of the quantity
$\rho_{\chi} \cdot \sigma^{\rm (nucleon)}_{\rm scalar}$ is performed,
values for the local density $\rho_{\chi}$ versus the relic abundance
$\Omega_{\chi}h^2$ may be deduced by proceeding in the following way
\cite{noi6}:

\noindent
1)  $\rho_{\chi}$ is evaluated as 
$[\rho_{\chi} \cdot \sigma^{\rm (nucleon)}_{\rm scalar}]_{expt}$ /
$\sigma^{\rm (nucleon)}_{\rm scalar}$, 
where $[\rho_{\chi} \cdot \sigma^{\rm (nucleon)}_{\rm scalar}]_{expt}$ 
denotes the experimental value, and 
$\sigma^{\rm (nucleon)}_{\rm  scalar}$ is calculated as indicated above;
2) to each value of  $\rho_{\chi}$ one associates the corresponding
calculated value of $\Omega_{\chi} h^2$. 
The scatter plot in Fig. \ref{fig:5} is derived from the lowest value of the
annual--modulation region of Ref.\cite{damalast},  
$[\rho_{\chi}$/(0.3 GeV cm$^{-3}$) $\cdot \sigma^{\rm (nucleon)}_{\rm
    scalar}]_{expt} = 1 \cdot 10^{-9}$ nbarn, and by taking 
$m_{\chi}$ in the range of Eq. (\ref{eq:mass}). 
 This  plot,
obtained in case of effMSSM, shows that the most interesting region,  
{\it i.e.} the one 
with  0.2 GeV cm$^{-3} \leq \rho_{\chi} \leq $ 0.7 GeV cm$^{-3}$ 
 and $0.05 \leq \Omega_{m} h^2 \leq 0.3$ (cross-hatched region in 
the figure), is covered
   by susy configurations probed by the WIMP direct detection.

Let us examine the various sectors of Fig. \ref{fig:5}. 
 Configurations above the upper horizontal line are
incompatible with the upper limit on the local density of dark
matter in our Galaxy and must be disregarded.
Configurations above the 
upper slanted dot--dashed line and below the upper horizontal solid line 
would imply a stronger clustering of neutralinos in our halo as 
compared to their average distribution in the Universe. This
situation may be considered unlikely, since in this case
neutralinos could fulfill the experimental range for 
$\rho_\chi$, but they would contribute only a small fraction to
the cosmological cold dark matter content.
For configurations which fall inside 
the band delimited by the slanted dot--dashed lines and simply--hatched 
in the figure,
the neutralino would provide only a fraction of the cold dark 
matter at the level of local density and of the 
average relic abundance, a situation which would be possible, for instance,
if the neutralino is not the unique cold dark matter particle
component. To neutralinos belonging to
 these configurations one 
should assign a {\it rescaled} local density (see Sect. 4.1.3). 

We remind that the scatter plot in Fig. \ref{fig:5} refers to a representative value of
$[\rho_{\chi}$ $\cdot \sigma^{\rm (nucleon)}_{\rm scalar}]$ {\it inside the
  current experimental sensitivity region}, thus the plot in Fig. \ref{fig:5} shows that
{\it current experiments of WIMP direct detection are probing relic neutralinos
  which may reach values of cosmological interest, but also neutralinos whose
  local and cosmological densities may provide only a very small fraction of
  these densities}. These properties were anticipated by the general arguments
previously discussed in Sect. IV.C.

\begin{figure}[t]
 \vspace{9.0cm}
\includegraphics{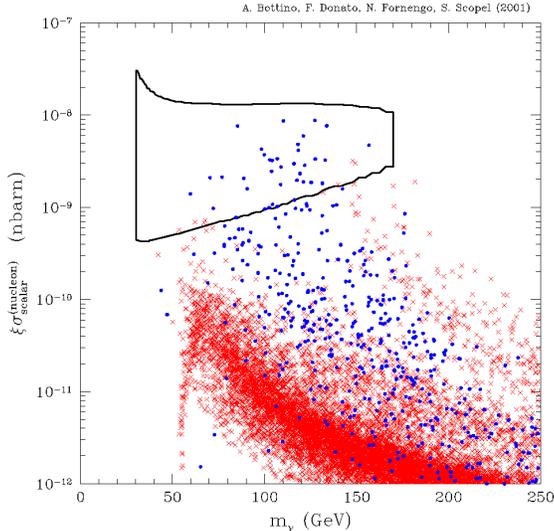}
 \caption{\it
Scatter plot of $\xi \sigma_{\rm scalar}^{(\rm nucleon)}$ versus 
$m_{\chi}$ in case of universal SUGRA. 
Set 1 for the quantities $m_{q}<\bar{q}q>$'s is employed \cite{noi6}.
Crosses (dots) denote 
configurations with $\Omega_{\chi} h^2 > 0.05$ 
($\Omega_{\chi} h^2 < 0.05$). The solid contour denotes the 3$\sigma$ annual--modulation
region of Ref.\cite{damalast} (with the specifications given in the text).
\label{fig:6a}}
\end{figure}

\begin{figure}[t]
 \vspace{9.0cm}
\includegraphics{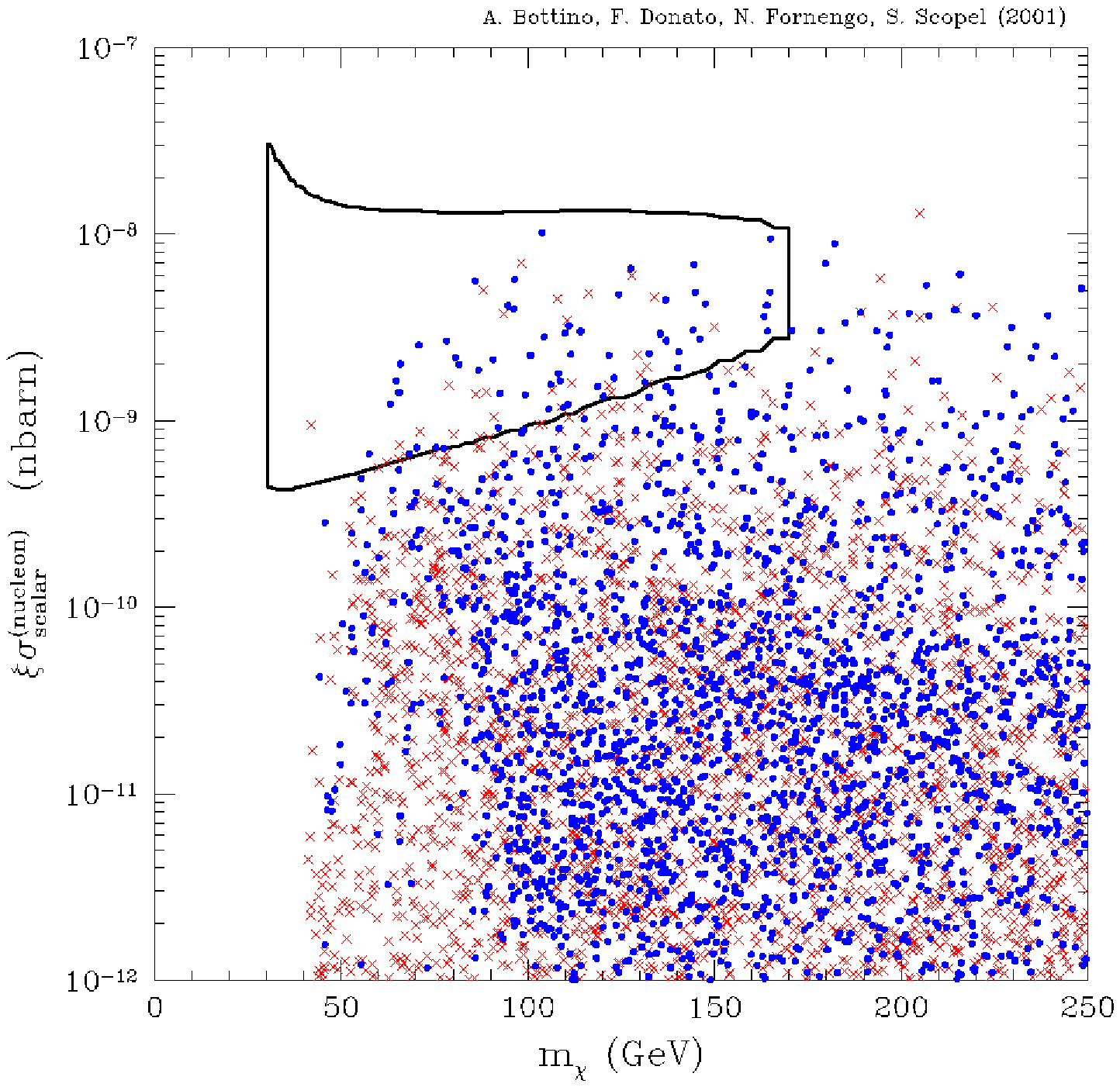}
 \caption{\it
Same as in Fig. \ref{fig:6a} in case of effMSSM.
\label{fig:6b}}
\end{figure}

For sake of comparison with specific experimental results, we provide in Figs.
\ref{fig:6a}-\ref{fig:6b} the scatter plots for the quantity $\xi \sigma^{\rm (nucleon)}_{\rm
  scalar}$ versus $m_{\chi}$ in the two supersymmetric schemes. In these
figures the solid line denotes the frontier of the 3$\sigma$ annual--modulation
region of Ref.\cite{damalast}, when only the uncertainties in $\rho_l$ and in
the dispersion velocity of a Maxwell--Boltzmann distribution, but not the ones
in other astrophysical quantities, are taken into account. As discussed in the
Introduction, effects due to a possible bulk rotation of the dark halo or to an
asymmetry in the WIMP velocity distribution would move this boundary towards
higher values of $m_{\chi}$.  Our results in Figs. \ref{fig:6a}-\ref{fig:6b} show that the susy
scatter plots reach up the annual--modulation region of Ref.\cite{damalast},
even with the current stringent bounds from accelerators (obviously, more
easily in effMSSM than in SUGRA).

Finally, we recall that use of set 2 for the quantities 
 $m_{q}<\bar{q}q>$'s instead of set 1 (see, for definitions, 
Ref.\cite{noi6}) would entail an increase of
 about a factor 3 in all the scatter plots of Figs. \ref{fig:6a}-\ref{fig:6b}.

\section{Conclusions}

In this work, after a short general introduction, 
 we have shown that the current direct experiments for
WIMPs, when interpreted in terms of relic neutralinos, are indeed
probing regions of the supersymmetric parameter space compatible with
all present bounds from accelerators. We have quantified the extent of
the exploration attainable by WIMP direct experiments in terms of
different  supersymmetric schemes, from a SUGRA scheme with unification
assumptions at the grand unification scale to an effective model,
effMSSM, at the electroweak scale. It has been stressed that, due the
large uncertainties in the unification assumptions in SUGRA schemes,
the effMSSM framework turns out to be the most convenient model for
neutralino phenomenology.
 
We have proved that part of the configurations probed by current WIMP
experiments entail relic neutralinos of cosmological interest, 
and,   {\it a fortiori},  
also neutralinos
which might contribute only partially to the required amount of dark
matter in the Universe. 
This last property was anticipated by the arguments presented 
in Sect. 4.3 and confirmed by our numerical results. 
 The
cosmological properties have been displayed in terms of a plot of the
local density versus the average relic abundance, {\it i.e.} in a
representation which proves particularly useful to summarize the
properties of relic neutralinos (see Fig. \ref{fig:5}). 

The question: {\it Are relic
  neutralinos of very low (local and cosmological) densities 
detectable by current experiments of WIMP direct detection ?}
finds a straightforward and affirmative answer 
 in the  $\rho_{\chi}$ versus $\Omega_{\chi} h^2$ plot.  
{\it Direct detectability is possible
  even for neutralino densities quite minuscule as compared to the
  ones of cosmological interest.}

In the present note, our discussions were mainly focussed on
implications of direct detection results. 
However, as mentioned in Sect. 4.2, also WIMP indirect searches are 
quite important. Measurements of up--going muon
fluxes from the center of the Earth and from the Sun 
can potentially  either find a signal or, at least, 
place significant constraints (though some uncertainties about a 
possible solar--bound population  have still  to be resolved)
 \cite{noiult}. Also measurements of low--energy antiprotons in space 
may  provide interesting  constraints on the susy model parameters
 \cite{noiult}, in
view of the new refined calculations of secondary antiprotons 
\cite{beu,bbe,donato}. 

We notice that, by the arguments presented in Sect. 4.3, similarly 
to the case of 
direct detection, the detectability of relic neutralinos by 
 measurements of up--going muon
fluxes from the center of the Earth and from the Sun at neutrino
telescopes is more favourable in case of low local and average
densities.

We have shown that the annual--modulation effect measured by the DAMA
Collaboration may be interpreted as due to relic neutralinos, which
are compatible with all current constraints from accelerator
measurements and WIMP indirect searches.  

In our evaluations we have taken into account that the determination
of the actual sensitivity region in terms of the WIMP--nucleon cross
section and of the WIMP mass from the experimental data depends quite
sizeably on uncertainties of various origins, mainly: i) possible
effects due to a halo bulk rotation and/or to asymmetries in the WIMP
velocities distribution, and  ii) significant uncertainties in the
determination of Higgs--quark--quark and neutralino--quark--squark
couplings. We stress that all these effects have to be taken
properly into account, when conclusions about comparison of theory
with experiments are drawn.

Finally, we wish to point out that  a susy Higgs boson at a mass of
about  115 GeV, as possibly hinted by the Higgs LEP experiments
\cite{higgs},  would fit remarkably well in the  scenario depicted
above \cite{el3}.

\section{Acknowledgements} 
This work was partially supported by the Research
Grants of the Italian Ministero dell'Universit\`a e della Ricerca
Scientifica e Tecnologica (MURST) within the {\sl Astroparticle
  Physics Project}. 
\vspace{2cm}

\end{document}